\RequirePackage[2020-02-02]{latexrelease}
\documentclass[aps,prb,superscriptaddress,10 pt,twocolumn]{revtex4-1}
\usepackage{graphicx}
\usepackage{dcolumn}
\usepackage{bm}
\usepackage{colordvi}
\usepackage{color}
\usepackage{cancel}

\begin{document}

\title{Spin flip locking by the tunneling and relaxation in a driven double quantum dot with spin-orbit coupling}

\author{D.V. Khomitsky}
\email{khomitsky@phys.unn.ru}
\affiliation{Department of Physics, National Research Lobachevsky State University of Nizhny Novgorod, 603022 Gagarin Avenue 23, Nizhny Novgorod, Russian Federation}
\author{M.V. Bastrakova}
\affiliation{Department of Physics, National Research Lobachevsky State University of Nizhny Novgorod, 603022 Gagarin Avenue 23, Nizhny Novgorod, Russian Federation}
\affiliation{Russian Quantum Center, 143025 Skolkovo, Moscow,  Russian Federation}
\author{D.S. Pashin}
\affiliation{Department of Physics, National Research Lobachevsky State University of Nizhny Novgorod, 603022 Gagarin Avenue 23, Nizhny Novgorod, Russian Federation}

\begin{abstract}
Coupled spin evolution and tunneling together with the relaxation and decoherence effects are studied for the double quantum dot formed in a semiconductor nanowire and driven by the periodic electric field. Such  structure represents a model of the spin and charge subsystems interacting via the strong spin-orbit coupling. It is found that at certain regimes the combination of fast relaxation in the coordinate channel with the slower relaxation in the spin channel leads to the promising combination of fast spin manipulation and slow spin relaxation, locking the flipped spin in an excited state in one of the dots for a sufficiently long time. The predicted effect is maintained for a wide range of the relaxation times and the driving amplitude both for the coordinate and the spin channels and is also observed on higher subharmonic which requires lower driving frequencies.  
\end{abstract}

\date{\today}
\maketitle

\section{Introduction}

The spin dynamics in low dimensional semiconductor structures demonstrates a great variety of both fundamental phenomena and promising applications in spintronics and quantum computations. The basic building block representing the qubit, i.e. the two-level system, has been the subject of intensive research where different phenomena such as the Landau-Zener-St\"uckelberg-Majorana (LZSM) interference \cite{Nori2010,Nori2023} have been studied. The coupled tunneling and the spin evolution in semiconductor structures with quantum dots have been the subject of intensive research
\cite{Semenov2007,Taylor2007,Platero2008,Petersson2010,Ludwig2013,Ribeiro2013,Vandersypen2013,Vandersypen2014,Loss2019, Krzywda2021,Platero2023} including the experimentally observed manifestation of the interplay between the tunneling and the spin flip associated with the electric dipole spin resonance (EDSR) and its subharmonics in the gate-defined single and double quantum dots (QD) in semiconductor structures \cite{Loss2002,Rashba2003,Loss2003,Debald2005,Loss2006,Yokoshi2009,Nowack2007,StehlikPetta2014,Studenikin2018,Studenikin2019,Studenikin2023}. 
Such systems may represent a model of the hybrid spin-charge qubit  encoded in the the spin and charge interacting subsystems \cite{Petta2006,Burkard2007,Zhang2007,Ramon2010,Petersson2010b,Mizuta2017,Yang2019,Kregar2020,Rancic2022}. Here the Zeeman splitting generates the two-level subspace of the spin subsystem and the double or multi-dot \cite{Platero2022} confinement potential provides a controlled charge distribution for the two-level charge (coordinate) subspace encoding the states by the electron location in left- or right-QD. The two subsystems may interact via the spin-orbit coupling intrinsically present in the given structure or generated by the non-uniform magnetic field \cite{PioroLadriere2007}  and driven by the monochromatic or two-photon Raman \cite{Bodey2019} mechanism. In a recent paper \cite{Zuk2024} the spin locking mechanism has been proposed as a tool for the dynamic decoupling in a system of the transmon qubits aimed on the reduction of the gate errors. Thus, it is of interest to consider further examples of the interacting spin and charge subsystems where the spin flip locking may take place prolonging the lifetime of the excited state.

Motivated by the mentioned experiments, we have studied in our recent papers \cite{KS2022,PRB2023} the LZSM effects for the coupled spin and charge dynamics driven in the double quantum dots by the periodic electric field and the spin-orbit coupling (SOC), focusing on the EDSR \cite{KS2022} and its subharmonics \cite{PRB2023}. It was found that the tunneling may enhance the SOC-induced spin flip via the interdot tunneling. The effects studied in these papers have been considered on relatively short time scales which allowed to neglect the relaxation effects and to consider only the coherent dynamics. 

When several, at least two interacting quantum dots are considered on the time interval comparable with the relaxation time in any of the subsystems, the enlarged size of the structure and the associated transport and tunneling effects inevitably call for the consideration of various relaxation and decoherence effects studied under the well-known density matrix formalism \cite{BP,Hanggi1997,GrifoniHanggi1998,Hone2009,Nalbach2009,Xu2014}. The applications include the already mentioned low dimensional semiconductor structures with quantum dots \cite{Semenov2007,Taylor2007,Platero2008,Petersson2010,Ludwig2013,Vandersypen2013,Vandersypen2013,Krzywda2021,Platero2023}, the qubits based on the Josephson junctions \cite{Makhlin2001,Satanin2012,Bastrakova2021,Pashin2023}, the circuit or cavity quantum electrodynamics \cite{Bonifacio2020,Tokman2023}, the entangled photons \cite{Marques2015}, the three-level systems \cite{Zhang2023}, and many others. The primary question is how well the results obtained within the coherent dynamics approximation survive on the specific time intervals and what qualitatively new effects the relaxation and decoherence may introduce into the coupled tunneling and spin dynamics in addition to the trivial damping and smearing of the valuable signals.

In this paper we include the relaxation and decoherence within the master equation approach for the electron density matrix and study the driven spin dynamics coupled with the interdot tunneling in a double quantum dot under the periodic driving by the electric field. Similarly to our previous studies \cite{KS2022,PRB2023}, we consider the resonant spin-flip transitions assisted by the interdot tunneling. It is found that, under some circumstances, the relaxation can play an unexpectedly stabilizing role for the spin flip, locking the flipped spin in the excited Zeeman state. The mechanism is activated in the resonant regime where the spin flip is accompanied by the interdot tunneling. The fast charge relaxation drives the electron into the single dot having the lower energy for the charge degree of freedom while the spin maintains its flipped state on the longer time scale which may go beyond the spin relaxation time due to the driving. This effect, similar to the ones observed in three-level atomic and laser systems \cite{Evers2005,Ilinova2012,Boukhris2023} but observed mainly in the $GHz$ frequency range in the semiconductor dot experiments \cite{Nowack2007,StehlikPetta2014,Studenikin2018,Studenikin2019,Studenikin2023}, can manifest itself for various system parameters and both on the main and the sub-harmonics of the resonance. It is of fundamental interest and is promising for the spin manipulation technique development in nanostructures. Our predictions may stimulate the development of coupled spin and charge qubit setups where the charge dynamics improves the characteristics of the spin evolution.

This paper is organized as follows. In Sec. II we describe the Hamiltonian and its lowest four-level subspace forming the coupled spin and charge  subsystems. The degree of the validity for the four-level model for the coherent dynamics is evaluated by searching for the borders in the parameter space which enclose the impact of the four lowest levels above the preset threshold. We extend the approach by introducing the evolution equation for the electron density matrix with the Lindblad operators and the relaxation rates for the charge and spin two-level subsystems, respectively. In Sec. III we present the numerical analysis of the regimes where the spin flip is assisted by the interdot tunneling and visualize the spin evolution on the associated Bloch sphere for each subsystem. We consider different types of resonances both without and with the interdot tunneling. In the regime with the interdot tunneling, we observe the unexpectedly stabilizing role of the relaxation at certain regimes on the spin flip in a single dot where the electron has been initialized in the ground state. Finally, in Sec. IV we give our conclusions.

\section{Model}

\subsection{The Hamiltonian}

The Hamiltonian which we use describes the single-particle states in a double quantum dot formed in a semiconductor nanowire by the gate potential. It has been derived and studied in detail in our preceding papers \cite{KS2022,PRB2023}. 
It is expressed via the sum

\begin{equation}
H(t)=H_0+V_d(t),
\label{ham}
\end{equation}
where the time-independent part of the Hamiltonian $H_0$ is

\begin{equation}
H_0=H_{2QD}+H_Z+H_{\rm{SO}}.
\label{ham0}
\end{equation}
In (\ref{ham0}) the contribution $H_{\rm{2QD}}$ corresponds to the movement of the electron in the double-minima potential of the quantum dot with the detuning, 

\begin{equation}
H_{\rm{2QD}}=k_x^2/2m+U_0(x)+U_d f_1(x). 
\label{h2qd}
\end{equation}
Here $m$ is the effective mass in the lowest subband of the size quantization in the nanowire (we use the units with $\hbar = 1$). The double-minima potential is modeled by the function $U_0(x)=U_0((x/d)^4-2(x/d)^2)$, where $2d$ is the interdot center distance and $U_0$ is the interdot barrier height. 
The last term $U_d f_1(x)$ in (\ref{h2qd}) describes the static detuning being the difference between the interdot minima. Here the function $f_1=(x/d_1)^3-3/2\cdot (x/d_1)^2$ models the smooth connection with the initial double well potential with $d_1=1.5 d$ and provides the bottom-down shift for the right QD at $U_d<0$. 

\begin{figure}[tbp]
\centering
\includegraphics[width=0.47\textwidth]{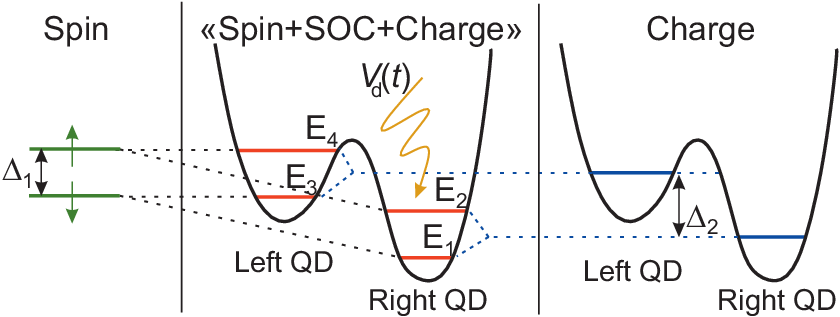}
\caption{Scheme of the spin (left) and charge (right) subsystems in a double quantum dot with the lowest subspace of the levels $E_1, \ldots, E_4$ of the coupled spin-charge system (center). The splitting $\Delta_1$ for the spin subsystem is generated by the Zeeman term (\ref{deltaz}), the splitting $\Delta_2$ for the charge subsystem is generated by the detuning which shifts the relative position of the dot energy minimum. The two subsystems are coupled via SOC and driven by the periodic electric field $V_d(t)$.} 
\label{figlevels}
\end{figure}

The second term in (\ref{ham0}) is the Zeeman coupling 

\begin{equation}
H_Z=\frac{1}{2}g\mu_B B_z \sigma_z 
\label{hz}
\end{equation}
generated by the static magnetic field which is the direction of the Oz axis, and $g$ is the effective $g$-factor, controlling the Zeeman splitting being the characteristic energy scale of the spin subsystem

\begin{equation}
\Delta_1=g\mu_B B_z. 
\label{deltaz}
\end{equation}
We may call the Zeeman-split pair of states $(\downarrow, \uparrow)$ in each dot as the spin subsystem with the distance between the levels $\Delta_1$ shown on the left panel of Fig.\ref{figlevels}. 

We will label the eigenstates and the energy levels of the coordinate part of the double dot Hamiltonian (\ref{h2qd}) as $\phi_n(x)$ and $\varepsilon_n$, respectively. The interdot minima splitting is the characteristic splitting $\Delta_2$ for the charge subsystem, 

\begin{equation}
\Delta_2=\varepsilon_{2}-\varepsilon_{1},
\label{vxtl0}
\end{equation}
where the numerical calculation of the energy levels gives us $\Delta_2=\gamma |U_d|$ with $\gamma \approx 0.55$.
For the negative detuning $U_d<0$ with $|U_d| \ll U_0$ the ground state $\varepsilon_1$ with the wavefunction $\phi_1(x)$ is located in the right QD with the exponentially small overlap with the left QD and the next state $\phi_2(x)$ is located in the left QD. This makes the basis for the charge two-level subsystem by attributing the wavefunction location to the right and left QD, respectively. In the following we will restrict ourselves to this two-level subspace for the coordinate degree of freedom shown schematically on the right panel in Fig.\ref{figlevels}.

The spin and charge subsystems are coupled via the third term in (\ref{ham0}) which is the SOC that may contain both Dresselhaus and Rashba terms. Here we will limit ourselves to the Dresselhaus term only since the Rashba term provides similar impact on the spin and charge evolution \cite{PRB2023}. For the leading order in the wavevector the SOC is linear for the GaAs-based nanowire, 

\begin{equation}
H_{\rm{SO}}=\beta_D \sigma_x k_x, 
\label{hso}
\end{equation}
where $\beta_D$ is the strength of the Dresselhaus term.
We perform the numerical diagonalization of the static Hamiltonian (\ref{ham0}),

\begin{equation}
H_0 \Psi_n=E_n \Psi_n, 
\label{sestatic}
\end{equation}
and obtain the set of energy levels $E_n$ and the two-component eigenfunctions $\Psi_n(x)$. 
The principal scheme for the lowest four-level subspace in (\ref{sestatic}) describing the two coupled spin and charge subsystems is given in the center panel of  Fig.\ref{figlevels} for the negative detuning $U_d<0$. In this basic example the level pairs ($E_1$, $E_2$) as well as ($E_3$, $E_4$) represent the spin subsystem with the splittings in each dot

\begin{equation}
E_{2}-E_{1}=\Delta^{(1)}_1, \quad E_{4}-E_{3}=\Delta^{(2)}_1,
\label{deltaline}
\end{equation}
which are slightly different from each other and from the Zeeman splitting $\Delta_1$ from (\ref{deltaz}) due to the corrections from the SOC term (\ref{hso}). 
The level pairs ($E_1$, $E_3$) and ($E_2$, $E_4$) belonging to the neighboring dots represent the charge subsystem with the splitting 

\begin{equation}
E_{3}-E_{1}=\Delta^{(1)}_2, \quad E_{4}-E_{2}=\Delta^{(2)}_2,
\label{entunnel}
\end{equation}
where again due to SOC the splittings $\Delta^{(1,2)}_2$ are slightly different from the spinless level splitting $\Delta_2$ defined in (\ref{vxtl0}).
It should be noted that due to the presence of the SOC the amplitude of the $z$-projection of the spin along the direction of the magnetic field for an eigenstate of the Hamiltonian (\ref{ham0}) is always less than $1$. It means that these eigenstates in (\ref{sestatic}) are not pure spin-up or spin-down states, although for the chosen parameters their in-plane spin projection is sufficiently smaller (about $1 \%$ in magnitude) than the $z$-projection. This allows us the consideration of the spin-split states in each QD as the states of the single spin subsystem similar to the pure Zeeman-splitted states $(\downarrow, \uparrow)$.

The second term in (\ref{ham}) is the periodic driving with the amplitude $V_d$, generating the time-periodic potential

\begin{equation}
V_d(t)=V_d \sin \omega t f_2(x).
\label{vxtg0}
\end{equation}
This potential describes the gate-defined electric field which is focused on the right QD only, modeled by the Gaussian shape $f_2(x)=\exp \left(-(x-d)^2/2d^2 \right)$. Below the system evolution is studied in the basis of the states (\ref{sestatic}) with the driving (\ref{vxtg0}).

\subsection{Four-level approximation}

First, we will take a look on the coherent driven dynamics without the relaxation in order to determine the regions in the parameter space where the four-level approximation is valid with the preset tolerance which is needful for the model of two coupled subsystems described above. The usual solution for the time-dependent Schr{\"o}dinger equation $i \partial \psi / \partial t= H \psi $ with the Hamiltonian (\ref{ham}) is found as a sum of the eigenfunctions from (\ref{sestatic}) with time-dependent coefficients \cite{KS2022,PRB2023}:

\begin{equation}
\psi(x,t) =\sum_n C_n(t) \Psi_n(x).
\label{psixt}
\end{equation}

\begin{figure}[tbp]
\centering
\includegraphics[width=0.48\textwidth]{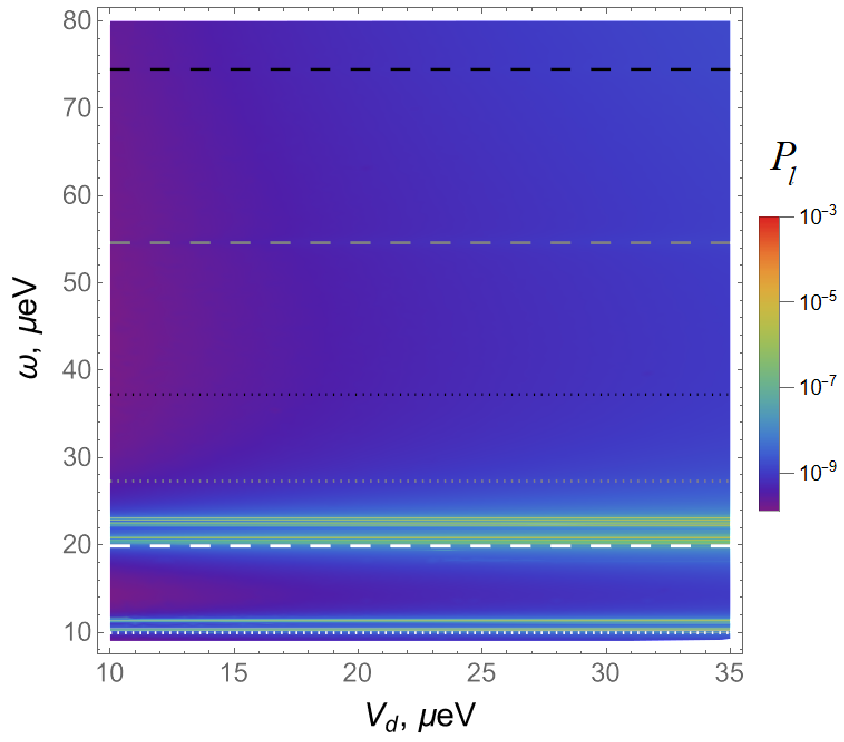}
\caption{Contour plots of the leakage probability (\ref{pleak}) in the plane of the driving field parameters $(V_d, \omega)$ describing the impact on the evolution produced by the higher levels above the lowest four-level subspace in Fig.\ref{figlevels} after $N=5000$ periods of the driving field. The black, grey and white dashed lines indicate the resonance transitions between the levels $E_1-E_4$, $E_1-E_3$ and $E_1-E_2$ in Fig.\ref{figlevels} for the splitting $\Delta_1=20 \mu eV$ and $\Delta_2=55 \mu eV$, and the dotted lines of the same colors indicate their second subharmonics. About $10^{-8} \ldots 10^{-7}$ of the leakage probability $P_l$ is observed along the line $\omega \sim 20 \mu eV$ during the hybridization of the $E_1-E_2$ level resonance and the high subharmonic of the $E_1-E_5$ resonance, as well for the other operating frequencies.}   
\label{figleak}
\end{figure}

To quantify the contribution from the higher levels above our basic four-level subspace, we will define the leakage probability to the higher levels

\begin{equation}
P_l=\max_{t} \left( 1-\sum_{i=1}^4 |C_i(t)|^2 \right)
\label{pleak}
\end{equation}
showing the maximum leakage over the whole evolution time. The most instructive application of (\ref{pleak}) is on the map of the parameters describing the driving field, i.e. on the $(V_d, \omega)$ plane. In Fig.\ref{figleak} we show the contour plots of (\ref{pleak}) after $N=5000$ periods of the driving field  $T=2\pi/\omega$ for the typical system parameters which include the electron effective mass in the lowest subband $m=0.11 m_0$ where $m_0$ is the free electron mass, the interdot minima distance $2d=116$ $nm$ and the interdot barrier $U_0=4$ $meV$ for the double dot potential in (\ref{h2qd}). The driving strength $V_d$ in Fig.\ref{figleak} varies between $10$ and $35$ $\mu eV$ which for the typical interdot barrier height $U_0=4$ $meV$ \cite{KS2022,PRB2023} gives the ratio $V_d/U_0=0.0025 \ldots 0.0087$, indicating that the interdot barrier shape is only weakly disturbed by the driving. The distance between the lowest four-level manifold and the next level $E_5$ (not shown in Fig.\ref{figlevels}) for such conditions is $\sim 2000 \mu eV$ which is an order of magnitude greater than the highest energy scale for the lowest four-level subspace and the driving in our model. The Dresselhaus amplitude in (\ref{hso}) is $\beta_D=3 meV \cdot nm$. The other system parameters are the following: the detuning $U_d=-100 \mu eV$ corresponding to the level splitting (\ref{vxtl0}) $\Delta_2$=$55 \mu eV$ and the Zeeman splitting $\Delta_1=20 \mu eV$. Such Zeeman splitting corresponds to the driving frequency $\sim 5$ $GHz$ which is within the frequency bands used in the experiments \cite{Studenikin2018}. The initial state for the evolution is the ground state $E_1$ in Fig.\ref{figlevels} which for the taken parameters is the spin-down state in the right QD. 

In Fig.\ref{figleak} the black, gray and white dashed lines indicate the resonance transitions between the levels $E_1-E_4$, $E_1-E_3$ and $E_1-E_2$ in Fig.\ref{figlevels} for the energy splitting parameters in (\ref{deltaz}) and (\ref{vxtl0}) taken as $\Delta_1=20 \mu eV$ and $\Delta_2=55 \mu eV$, and the dotted lines of the same colors indicate their second subharmonics. From Fig.\ref{figleak} one can observe the leakage probability of $10^{-8} \ldots 10^{-7}$ for all operating frequencies in our modeling which provides a good justification of working within the lowest four-level subspace for the most parts of the parameter map. 

A remarkable feature in Fig.\ref{figleak} is that the leakage probability is enhanced at certain frequencies, i.e. on some horizontal lines in Fig.\ref{figleak} where and obvious increase is seen also with the growing driving amplitude. These frequencies, grouped at $\omega \sim 20 \mu eV$ being the spin subsystem splitting (\ref{deltaline}), correspond to the hybridization of the $E_1-E_2$ level resonance and a high ($k \sim 100$) subharmonic of the $E_1-E_5$ resonance with the interlevel distance $E_5-E_1 \sim 2000 \mu eV$. Such amplification of the resonance at the hybridization points in the parameter space has been observed in our preceding paper \cite{PRB2023}. We can draw a conclusion that operating on the basic EDSR frequency $\omega=E_2-E_1$ seems to be not the best option due to the intensified leakage to the higher levels above the four-level manifold. A better way of flipping the spin can be the operations on the $E_1-E_4$ resonance or its subharmonic labeled by the black dashed and dotted lines in Fig.\ref{figleak}, respectively. We will gather more justifications on this choice in the next Section.

\subsection{Two-subsystem representation}

As it can be seen from Fig.\ref{figleak}, most of the driving field parameter plane provides a good background for the modeling the evolution within the four-level ground subspace  $E_1, \ldots, E_4$ illustrated by Fig.\ref{figlevels} which is spanned by the wavefunctions $\Psi_1, \ldots \Psi_4$ obtained from (\ref{sestatic}). We will consider them as the states for the two-subsystem representation of our system, the first one being the spin subsystem and the second one is the charge subsystem. The two subsystems interact via the SOC (\ref{hso}) and are both driven by the periodic field (\ref{vxtg0}). We would like to stress that in our double dot system both the spin and charge subsystems are not localized within a single dot, as it is routinely expected. The double dot setup serves as a source of a four-level manifold describing the ensemble of two coupled dots. The spin and charge subsystems can be extracted after the representation of the time-dependent Hamiltonian (\ref{ham}) for the two interacting parts is made in terms of the Pauli matrices  $\sigma_k^{(q)}$, $k=0, \ldots, 3$, attributed to each of the spin ($q=1$) and charge ($q=2$) subsystems:

\begin{eqnarray}
H_{4\times4}=c_{00} \cdot\sigma^{(1)}_0\otimes\sigma^{(2)}_0+c_{03} \cdot\sigma^{(1)}_0\otimes\sigma^{(2)}_3+
\nonumber
\\
+c_{30} \cdot\sigma^{(1)}_3\otimes\sigma^{(2)}_0+c_{33} \cdot\sigma^{(1)}_3\otimes\sigma^{(2)}_3+
\nonumber
\\
+F_{4\times 4} V_d \sin(\omega t).
\label{h4x4}
\end{eqnarray}
In (\ref{h4x4}) the coefficients $c_{ij}$ are expressed via the energies $E_n$ from the stationary Schr{\"o}dinger equation (\ref{sestatic}) as

\begin{eqnarray}
c_{00}&=&\frac{E_1+E_2+E_3+E_4}{4}, \nonumber \\
c_{03}&=&\frac{-E_1-E_2+E_3+E_4}{4}, \nonumber \\
c_{30}&=&\frac{-E_1+E_2-E_3+E_4}{4}, \nonumber \\ 
c_{33}&=&\frac{E_1-E_2-E_3+E_4}{4}. 
\label{cij}
\end{eqnarray}
The first coefficient $c_{00}$ corresponds to the shift of the energy zero and does not affect the system dynamics. The coefficient $2 c_{03} = (\Delta^{(1)}_2+\Delta^{(2)}_2)/2$ is the averaged distance from (\ref{entunnel}) between the levels of the charge subsystem. The coefficient $2 c_{30}=(\Delta^{(1)}_1+\Delta^{(2)}_1)/2$ is the averaged Zeeman splitting from (\ref{deltaline}) slightly corrected by the presence of SOC. The last coefficient $c_{33}$ is zero for the case of the equal Zeeman splittings in the left and right QDs and its nonzero value reflects the slight difference between the spin splitting in two dots created by the combination of SOC and detuning. This coefficient reflects the SOC-induced interaction between the spin and charge subsystems and in our modeling has a typical amplitude $|c_{33}| \sim 10^{-4}$ $\mu eV$ which allows treating SOC as a perturbation while keeping the concept of the spin and charge two-level subsystems, each described by its own relaxation parameters studied in the next Subsection.
The last term $F_{4\times 4}$ in (\ref{h4x4}) is given by the decomposition of the operator $I^{(1)}_{2\times 2}\otimes f_2(x)$, where $I^{(1)}_{2\times 2}$ is $2\times 2$ identity matrix in the subspace of the spin subsystem, and essentially depends on the shape of the driving function $f_2(x)$, creating in general the matrix $F_{4\times4}$ with all non-zero elements. By analyzing the representation (\ref{h4x4}) one can see that the spin subsystem is described by the third term with the level splitting given by $c_{30}$ while the charge subsystem is described by the second term in (\ref{h4x4}) with the level splitting given by the $c_{03}$. Both the spin and charge subsystems in the representation (\ref{h4x4}) do not belong to a single dot but to the ground four-level manifold of the whole double dot structure. After being introduced, the notation (\ref{h4x4}) allows us to apply the standard technique of the density matrix formalism for the evolution of the system of two coupled  subsystems described in the next Section.

\subsection{Evolution of the density matrix}

Since we are interested in the evolution modeling on different time scales including the ones comparable with the relaxation times, we will include the relaxation and decoherence for our system via the density matrix approach. We suppose that each of the spin and charge subsystems interacts with its corresponding thermostat independently and is described by the specific phase and energy dissipators and the relaxation rates $\Gamma_\varphi^{(q)}$, $\Gamma_e^{(q)}$ for the spin and charge subsystems, providing the additive impacts into the global system dissipator \cite{Hofer2017,Cattaneo2019}. The full two-subsystem density matrix describes the evolution of two coupled subsystems governed by the equation written in the Lindblad form in the instant (adiabatic) basis \cite{Nori2023,Makhlin2001,Bastrakova2021}: 

\begin{eqnarray}
\label{Lind}
\frac{\partial\rho}{\partial t} &=& -i [H_{4\times4},\rho] +\Gamma \rho, \nonumber \\ 
\Gamma &\equiv& \sum_{q=1}^2 \left( \frac{\Gamma_\varphi^{(q)}}{2} D[\overline{\sigma}^{(q)}_3]+\Gamma_e^{(q)} D[\overline{\sigma}^{(q)}_-]
\right),  
\end{eqnarray}
where the dash over the Pauli matrices indicates that they are written in the adiabatic basis. In Eq.(\ref{Lind}) the Hamiltonian  $H_{4\times4}$ from (\ref{h4x4}) is described in the previous Subsection, and the Lindblad operators are given by

\begin{eqnarray}
D[a] \rho &\equiv&   a \rho a^\dagger  - \frac{1}{2}\{a^\dagger a,\rho\}, \nonumber \\ 
\overline{\sigma}^{(q)}_- &\equiv& \frac{1}{2} (\overline{\sigma}^{(q)}_1 - i \overline{\sigma}^{(q)}_2).
\label{Diss}
\end{eqnarray}

We consider the approximation where each of the spin and charge subsystems interacts with its own dominating source of relaxation. For the GaAs double dots it can be the interaction with the nuclear spins for the spin subsystem and the interaction with the structural defects and phonons for the charge subsystem. This approximation allows treating their reservoir parameters separately, introducing the phase and energy relaxation rates $\Gamma_\varphi^{(q)}$ and $\Gamma_e^{(q)}$ for each subsystem ($q=1,2$) as follows \cite{Makhlin2001}:

\begin{eqnarray}
\label{gamma}
\Gamma_\varphi^{(q)}&=&\pi \alpha^{(q)} 2 k_B \tau \cos^2 \eta^{(q)},\nonumber \\ 
\Gamma_e^{(q)}&=&\pi \alpha^{(q)} \Delta \overline{E}^{(q)} \coth{\frac{\Delta \overline{E}^{(q)}}{2 k_B \tau }} \sin^2 \eta^{(q)}.
\end{eqnarray}

It should be mentioned that the energy relaxation rates in Eq.(\ref{gamma}) are energy-dependent, i.e. they vary with the inter-level distance $\Delta \overline{E}^{(q)}$ for the adiabatic basis. In Eq.(\ref{gamma}) $k_B$ and $\tau$ are the Boltzmann constant and the thermostat temperature, and the parameters $\eta^{(q)}$ and $\alpha^{(q)}$ describe the subsystem-thermostat coupling type and strength, respectively. Despite the time dependence of the rates (\ref{gamma}), we need a parameter reference for the relaxation times for the spin and coordinate subsystem. To introduce this, we will use the characteristic relaxation times defined via (\ref{gamma}) at the initial moment $t=0$:  $T_\varphi^{(q)}=2\pi/\Gamma_\varphi^{(q)}(0)$ and $T_e^{(q)}=2\pi/\Gamma_e^{(q)}(0)$. The energy parameters $\Delta \overline{E}^{(q)}$ describe the inter-level distance in the instant basis, 

\begin{eqnarray}
\label{et}
\Delta \overline{E}^{(1)}=2\overline{c}_{30}(t), \nonumber \\ 
\Delta \overline{E}^{(2)}=2\overline{c}_{03}(t),
\end{eqnarray}
where $\overline{c}_{ij}(t)$ can be found from (\ref{cij}) 
by the formal replacement of the energies $E_i$ for the stationary part of the Hamiltonian in (\ref{sestatic}) by the instant basis energies $E_i (t)$ which can be found from the equation

\begin{eqnarray}
\label{time_ind_sсh}
H(t) \Psi_i(t) =  E_i(t) \Psi_i(t), 
\end{eqnarray}
where $H(t)$ is the time-dependent Hamiltonian (\ref{ham}).
It should be mentioned that the validity of Eq.(\ref{Lind}) is justified if the adiabaticity condition \cite{Lidar2016} is satisfied: 

\begin{equation}
h/\delta^2 \ll 1, 
\label{adiabcond}
\end{equation}
where

\begin{eqnarray}
\delta=\displaystyle\min_{t, n, m}\left|E_n(t)-E_m(t)\right | \quad{\rm and} \\   
h=\displaystyle\max_{t, n, m}\left| \left \langle \psi_{n} (t)\left | \partial_t H (t) \right | \psi_{m} (t)\right \rangle \right |.
\label{deltah}
\end{eqnarray}
Our calculations show that the typical value of the adiabaticity parameter $h/\delta^2$ in (\ref{adiabcond}) is $\sim 0.02 \ldots 0.07$ which justifies the evolution equation in the form of Eq.(\ref{Lind}).

After solving Eq.(\ref{Lind}) for the given parameters of the subsystems and the driving in Eq.(\ref{h4x4}) we calculate the average values tracking the evolution of each subsystem: the spin projections $\langle \sigma_{x,y,z} \rangle (t)$ tracking the state of the spin subsystem where we are mainly interested in the evolution of the $z$-projection of the spin, and the electron position $\langle x/d \rangle (t)$ tracking the state of the charge subsystem.

\section{Numerical results}

\subsection{EDSR affected by the spin relaxation}

\begin{figure}[tbp]
\centering
\includegraphics[width=0.47\textwidth]{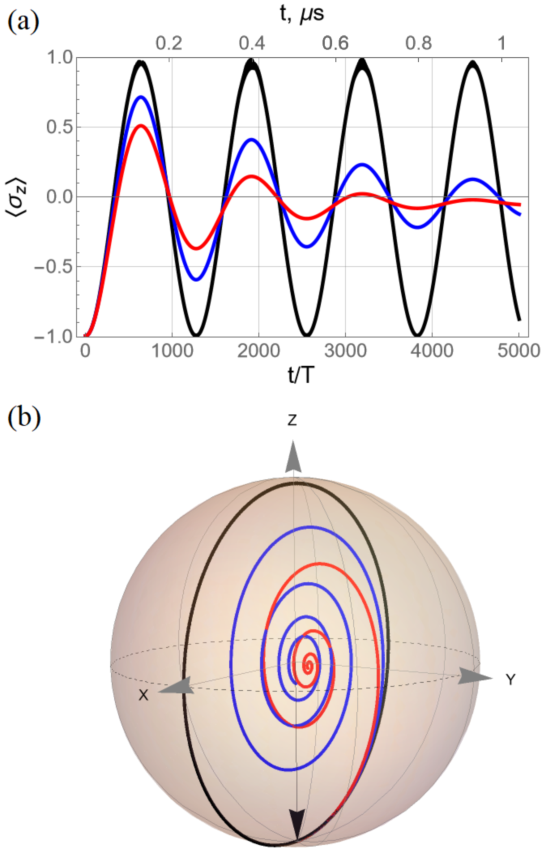}
\caption{(a) Stroboscopic evolution of $\langle \sigma_z \rangle (t)$ (in units of $\hbar/2$) for the main EDSR harmonic at the resonance between the levels $E_1$ and $E_2$ in Fig.\ref{figlevels}. Plot is shown vs time (upper scale) and vs the number of driving field periods (lower scale); (b) Stroboscopic evolution of the vector (\ref{spinvector}) for the spin subsystem ($q=1$) shown on its Bloch sphere. The system parameters are the same as in Fig.\ref{figleak} with the driving frequency defined by the Zeeman splitting $\Delta_1=20 \mu eV$ and the amplitude $V_d=30$ $\mu eV$. The initial state is the ground spin-down state with energy $E_1$ in the right QD marked by the black downward arrow. Black curve is for the absence of coupling with the thermostat. Blue curve is for the relaxation times $T_\varphi^{(2)}=T_e^{(2)}=0.4$ $\mu s$, and we take everywhere $T_\varphi^{(1)}=T_e^{(1)}=10T_\varphi^{(2)}$. Red curve is for $T_\varphi^{(2)}=T_e^{(2)}=0.2~\mu s$. The thermostat temperature in two latter cases $\tau=100$ mK. 
The relaxation provides trivial damping for pure two-level spin dynamics without tunneling.} 
\label{figmain12}
\end{figure}

We consider a typical double dot formed by electrostatic gates in a GaAs nanowire similar to those used in the experiments \cite{Studenikin2018} and modeled in our preceding papers \cite{KS2022,PRB2023}.
We begin with a discussion of the relaxation effects on the most basically considered pure EDSR regime.
In Fig.\ref{figmain12} we show the stroboscopic evolution of the $z$-projection of the spin $\langle \sigma_z \rangle (t)$ (in units of $\hbar/2$) for the moments of time $t=nT$ where $T$ is the period of the driving field obtained after solving Eq.(\ref{Lind}) on $5000$ $T$ time interval with the frequency matching the level distance $E_2-E_1$ (see Fig.\ref{figlevels}) defined by the Zeeman splitting $\Delta_1=20 \mu eV$. The total evolution time is about $1$ $\mu s$. 
The system parameters are the same as in Subsection II.B and in Fig.\ref{figleak} where we choose the moderate driving amplitude $V_d=30$ $\mu eV$.
On panel (a) the plot is shown vs time measured in microseconds (upper scale) in the driving field period (lower scale). On panel (b) the stroboscopic evolution of all three spin projections representing the spin subsystem is shown on its Bloch sphere which can be introduced for both spin and charge subsystem separately,

\begin{equation}
{\bf S}^{(q)}(t)=\left( \langle \sigma_1^{(q)} \rangle (t), \langle \sigma_2^{(q)} \rangle (t), \langle \sigma_3^{(q)} \rangle (t) \right). 
\label{spinvector}
\end{equation}
Here the spin and charge subsystem evolution corresponds to $q=1$ and $q=2$ in (\ref{spinvector}), respectively. The initial state is the ground spin-down state with energy $E_1$ in the right QD (see Fig.\ref{figlevels}), being on the south pole of the sphere in Fig.\ref{figmain12}(b). Black curve shows the coherent evolution when there is no coupling with the thermostat. Blue curve is for the relaxation times $T_\varphi^{(2)}=T_e^{(2)}=0.4~\mu s$. As to the spin relaxation, we consider it to be much longer, $T_\varphi^{(1)}=T_e^{(1)}=10T_\varphi^{(2)}$. Red curve is for the shorter relaxation times $T_\varphi^{(2)}=T_e^{(2)}=0.2~\mu s$, and again, $T_\varphi^{(1)}=T_e^{(1)}=10T_\varphi^{(2)}$. The thermostat temperature for the blue and red curves is $\tau=100$ $mK$. We have considered rather long relaxation times in order to illustrate the evolution on the smaller EDSR frequency in this Subsection (compared to the frequencies in the next Subsection) before they are damped completely.  

The analysis of the evolution in Fig.\ref{figmain12} indicates that the relaxation provides just the simple damping for the pure two-level spin dynamics at the EDSR if the tunneling is effectively turned off. Several incomplete spin rotations are possible, but the onset of the spin relaxation dampers the spin oscillations after several $\mu s$, leading to the need of stronger driving which will reduce the probability of the system to stay within the lowest four-level subspace of the double dot. 
Thus, we may conclude that another type of the resonance, maybe involving the tunneling in addition of the spin flip, can be of interest for the efficient spin manipulation, which is described in the following Subsection.

\subsection{Spin flip locking by the tunneling and relaxation}

We continue with the evolution analysis for another type of resonance for the pair of levels $E_1$ and $E_4$ in Fig.\ref{figlevels} where the spin flip is accompanied by the effective interdot tunneling\cite{KS2022,PRB2023}. Here the basic system parameters are the same as in the previous Subsection. The frequency of the resonance between the levels $E_1$ and $E_4$ is determined by the sum $\Delta_1+\Delta_2=75$ $\mu eV$ giving us the linear frequency  $f \sim 18~GHz$. It is still within the frequency bands used in the experiments \cite{Studenikin2018}. The characteristic timescale for the developing spin and coordinate dynamics is much faster than in the previous Subsection. It is in between of the charge and spin relaxation timescales which allows to capture well-established spin rotations.  
One should mention that the amplitude of the driving field $V_d=30 \mu eV$ is lower than the detuning amplitude, meaning that we are not in the effective LZSM regime with the level anticrossing \cite{Nori2023,KS2022} but purely in the resonance at $E_4-E_1=\omega$. The initial state for the evolution is again the ground state $E_1$. The total evolution time for the modeling is about 100 $ns$ which is within the range of the earlier experiments on EDSR-induced current oscillations in GaAs gate-defined double dots where no significant damping has been observed \cite{Nowack2007}. 

\begin{figure}[tbp]
\centering
\includegraphics[width=0.49\textwidth]{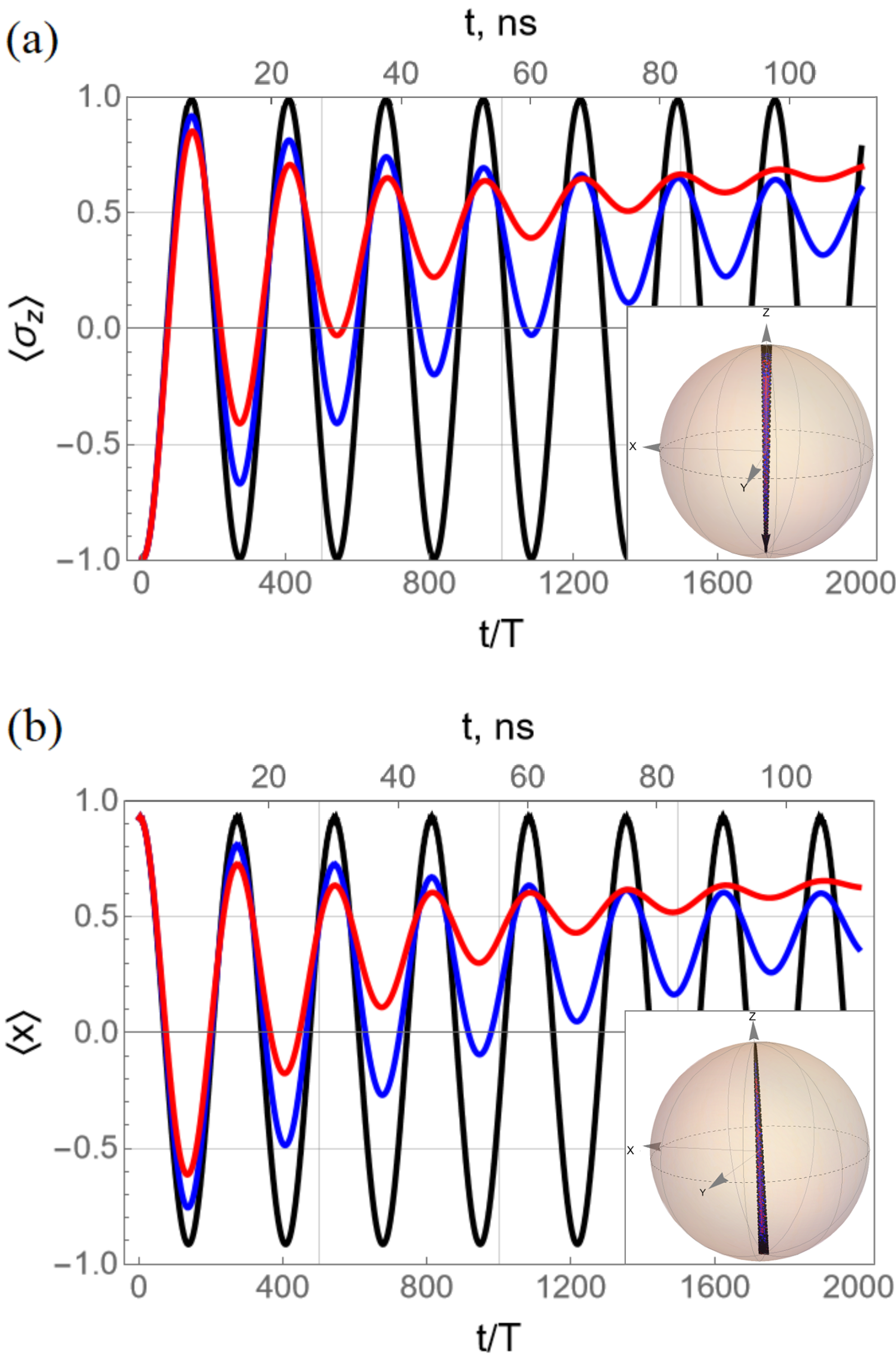}
\caption{Evolution of (a) the mean spin projection $\langle \sigma_z \rangle (t)$ and (b) the mean coordinate $\langle x \rangle (t)$ (in units of $d$) for the spin-flip tunneling resonance between the levels $E_1$ and $E_4$ on the frequency close to the combined splitting $\Delta_1+\Delta_2=75$ $\mu eV$ and accompanied by the quick relaxation between the levels $E_4$ and $E_2$ in Fig.\ref{figlevels}. The evolution is plotted vs time (upper scale) and the number of driving field periods (lower scale). All the other parameters and the labeling of the curves are the same as in Fig.\ref{figmain12}. The stabilization of the flipped spin in the right QD can be observed due to the fast charge relaxation and slow spin relaxation. The insets show the evolution of the vector (\ref{spinvector}) on the Bloch sphere for (a) the spin subsystem and (b) the charge subsystem. The initial state is on the south pole for both subsystems.} 
\label{figshort14}
\end{figure}

In Fig.\ref{figshort14} we show the evolution for the mean values of the spin projection $\langle \sigma_z \rangle (t)$ and the coordinate $\langle x \rangle (t)$, the latter measured in units of $d$, obtained after solving Eq.(\ref{Lind}) on the $2000$ $T$ with the frequency matching the level distance $E_4-E_1$ (see Fig.\ref{figlevels}). The associated evolution on the Bloch sphere of the vector (\ref{spinvector}) is shown in the insets on Fig.\ref{figshort14} for (a) spin subsystem and (b) charge subsystem. The initial state is on the south pole for both subsystems. It is clear that for all considered values of the relaxation parameters the evolution on the Bloch sphere is similar for both subsystems and can be described as a small-radius rotation around the $z$ axis with the full-scale oscillations between the south and the north poles. The final position of the vector (\ref{spinvector}) in the insets of Fig.\ref{figshort14}, as it can be concluded from the plots in Fig.\ref{figshort14}, is on the north pole for the spin subsystem in panel (a) and on the south pole for the charge subsystem in panel (b).

As in the previous Subsection, we consider three sets of the relaxation parameters. For the first set corresponding to the black curves there is no coupling with the thermostat, i.e. it is the basic case of the coherent evolution studied earlier \cite{KS2022,PRB2023}. For the second set corresponding to the blue curves the charge subsystem relaxation times $T_\varphi^{(2)}=T_e^{(2)}=0.4$ $\mu s$, and the spin subsystem relaxation times are ten times slower, $T_\varphi^{(1)}=T_e^{(1)}=10T_\varphi^{(2)}$. This is a typical situation in semiconductor nanowires and quantum dots where the spin relaxation is substantially slower than the momentum relaxation. The spin relaxation time in the GaAs-based double dot system with the detuning is expected to be much longer, at least an order of magnitude \cite{Vandersypen2013,Vandersypen2014}. For the third set described by the red curves the relaxation is faster, $T_\varphi^{(2)}=T_e^{(2)}=0.2$ $\mu s$, and $T_\varphi^{(1)}=T_e^{(1)}=10T_\varphi^{(2)}$. The fastest charge relaxation time of $200$ $ns$ is of the same order as the maximum evolution time of $700$ $ns$ in Fig.\ref{figshort14} and is comparable to the coherence time obtained in recent calculations for the double dots with detuning \cite{Krzywda2021}. The considered scale of spin and charge relaxation times means that the charge (coordinate) subsystem evolves during the evolution towards its ground state which is the lowest level $E_1$ in the right QD while the spin subsystem is still far from its ground state. The thermostat temperature for the second and third cases is set at $\tau=100$ $mK$ being in agreement with the typical experimental conditions \cite{Studenikin2018}. Such temperature corresponds to the thermal energy of about $8.75 \mu eV$ being lower than all the other energy parameters of the model, meaning that the thermal-induced jumps to the higher levels have low probability. 

\begin{figure}[tbp]
\centering
\includegraphics[width=0.49\textwidth]{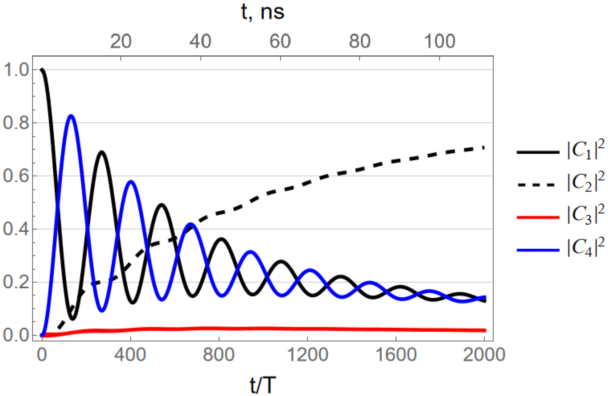}
\caption{An example of the level occupancies $|C_i(t)|^2$ evolution for the four-level subspace in Fig.\ref{figlevels} is shown for the case of strong relaxation labeled by the red curves in Fig.\ref{figshort14}. The growing behavior of $|C_2(t)|^2$ describing the steady flipped spin in the right QD is explained as the combination of the resonant dynamics between the levels $E_1$ and $E_4$ accompanied by the fast charge relaxation between the levels $E_4$ and $E_2$.} 
\label{figevolc}
\end{figure}

The most striking feature in Fig.\ref{figshort14} is the observed spin flip stabilization, or spin flip locking in the right QD when the relaxation is introduced in combination with the interdot tunneling. This effect can be explained by comparing the fast charge relaxation with the slow spin relaxation and the resonant driving. Namely, the driving on the resonance $E_4-E_1=\omega$ excites the spin-flip transitions accompanied by the interdot tunneling from the right to the left QD. 
This is illustrated in Fig.\ref{figevolc} where an example of the level occupancies $|C_n(t)|^2$ for the levels $E_1, \ldots, E_4$ from Fig.\ref{figlevels} is shown for the case of strong relaxation labeled by the red curve in Fig.\ref{figshort14}. The main purpose of Fig.\ref{figevolc} is to show the evolution in the basis of the coordinate- and spin-resolved states of the static Hamiltonian (\ref{ham0}) which is readily connected to the evolution of the mean values in Fig.\ref{figshort14}. In Fig.\ref{figevolc} one can see that for the resonance between the levels $E_1$ and $E_4$ the off-resonance spin-up level $E_2$ occupation $|C_2(t)|^2$ monotonically grows with time until a certain saturation of the spin flip in the right QD is achieved.
The populations $|C_1(t)|^2$ and $|C_4(t)|^2$ oscillate with the associated Rabi frequency for this two-level resonance modulated by the decaying envelope function while the spin-down level $E_3$ population $|C_3(t)|^2$ in the left QD demonstrates the slow-amplitude steady behavior. The observed Rabi frequency is higher for the $E_1$ - $E_4$ resonance than for the EDSR resonance considered in the previous Subsection which has been observed in our previous papers \cite{KS2022,PRB2023} since the former involves the greater spatial interdot displacement leading to a higher matrix element amplitude for the SOC. As to the stable and slow growth of $|C_2(t)|^2$, it is a result of the two processes: the fast interdot spin flip tunneling and the slow interdot charge relaxation. As a result, from Fig.\ref{figevolc} it is evident that a steady state in terms of the level populations is formed on longer times with the stable spin flip in the right QD.
We would like to stress that the considered example of the dynamics in Fig.\ref{figshort14} - Fig.\ref{figevolc} demonstrates how useful the second (charge) subsystem can be for the creation of the locked population of the excited state of the first (spin) subsystem, finally leaving the second subsystem in its ground state in the right QD.

We believe that the population behavior in Fig.\ref{figevolc} and the associated spin and coordinate evolution in Fig.\ref{figshort14} can be explained as the combination of the resonant dynamics between the levels $E_1$ and $E_4$ accompanied by the fast charge relaxation between the levels $E_4$ and $E_2$. Indeed, from the upper $E_4$ level the fastest relaxation is the charge one, stimulating the transition to the level $E_2$ in the right QD but leaving the same flipped spin which holds for the sufficiently long time as it can be seen in Fig.\ref{figshort14} and Fig.\ref{figevolc}. At longer times, when the spin relaxation stimulates the backward spin flip during the the transition to the ground level $E_1$, the resonant driving field will trigger the described process once again. As a result, one can expect a quite long lifetime of the flipped spin in the same right QD where the electron has been initialized. Such steady state formation resulting into the spin polarization has been observed, for example, for the nuclear spins in double quantum dots \cite{Petta2008,Schuetz2014}, for the optically driven electron spin coupled to the bath of nuclear spins \cite{Vezvaee2021}, for the driven spin coupled to a bath of ultracold fermions \cite{Knap2013}, and for the dynamics of two coupled qubits in the presence of the noise \cite{DasSarma2016}. This tunneling and relaxation- stabilized spin flip mechanism can be promising for spintronic applications.

\begin{figure}[tbp]
\centering
\includegraphics[width=0.47\textwidth]{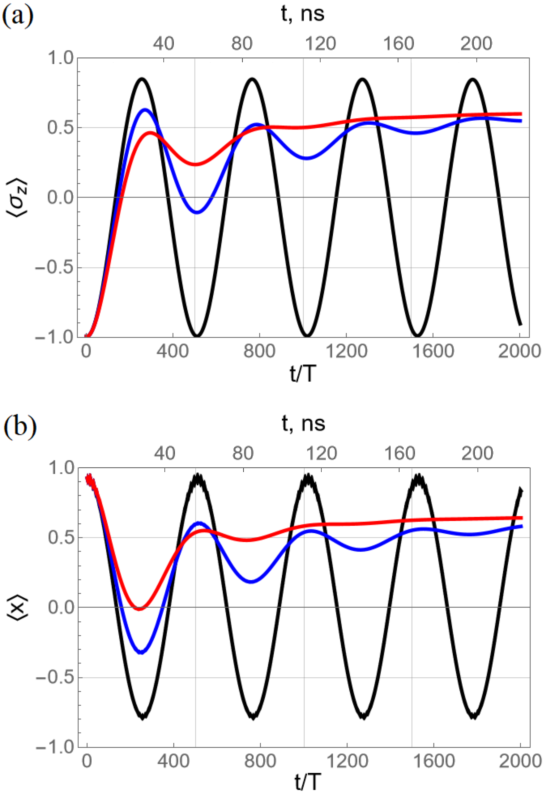}
\caption{The same as in Fig.\ref{figshort14} for the second subharmonic $s=2$ in (\ref{subharm}) of the $E_1$ - $E_4$ resonant transition in Fig.\ref{figlevels}. The labeling of the curves is the same as in Fig.\ref{figmain12} and Fig.\ref{figshort14}. The type of the resonance and the parameters (except the lower driving frequency and longer evolution time) are the same as in Fig.\ref{figshort14}. The relaxation leads to the dampening of the spin and coordinate oscillations and to the stabilization of the flipped spin.} 
\label{fig14sh2}
\end{figure}

\subsection{Evolution on the second subharmonic}

The relaxation effects become stronger for the EDSR subharmonics which frequency satisfies the condition 

\begin{equation}
\omega_s=\frac{E_4-E_1}{s}, \quad s=2,3, \ldots. 
\label{subharm}
\end{equation}
The reason is that the driving field on the subharmonics operates on lower frequencies and, as a result, the spin flip is achieved on longer times, being stronger affected by the relaxation. The driving frequency for the main harmonic of the $E_1$ - $E_4$ resonance for the considered Zeeman coupling and detuning is in the $18$ $GHz$-range which may not be easily accessible. Hence, the driving on one of the the subharmonics (\ref{subharm}) can be a viable option. We will consider the closest subharmonic $s=2$. In Fig.\ref{fig14sh2} we show the spin and coordinate evolution for the second subharmonic $s=2$ on the timescale of $\sim 250 ~ns$ covering $2000$ $T$.
All other parameters are the same as in Fig.\ref{figshort14}. As for the main harmonic, one can see that on the subharmonic the relaxation leads to the dampening of the spin and coordinate oscillations and to the stabilization of the flipped spin. An interesting and potentially useful feature is that the stabilized flipped spin projection is somewhat higher for the case with the stronger relaxation (red curve vs the blue curve). We attribute this difference to the progressing effects of the charge and spin relaxation which come into play more quickly (measured in the driving field period) on the subharmonics due to the lower driving frequency, leading to the locking of the electron in the lower state $E_2$ in the right QD with the flipped spin, compared to the higher state $E_4$ in the left QD. The predicted spin flip locking can be considered as an unexpectedly positive effect of the relaxation, maintaining the spin in the flipped state for sufficiently long time.

\subsection{Spin flip amplitude dependence on the parameters}

\begin{figure}[tbp]
\centering
\includegraphics[width=0.49\textwidth]{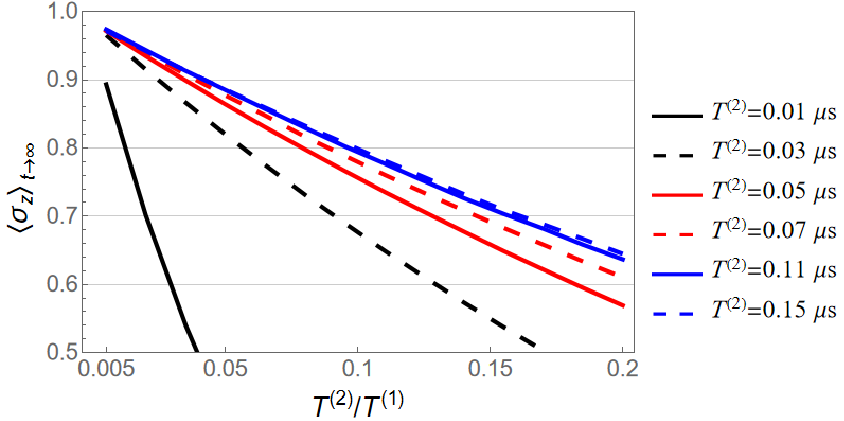}
\caption{Stabilized value of the spin projection $\langle \sigma_z \rangle_{t \to \infty}$ for different charge relaxation times $T^{(2)}$ labeled as differently shaped curves, each of them plotted vs the ratio $T^{(2)}/T^{(1)}$ of the charge and spin relaxation time. The other parameters are for the main harmonic of the $E_1-E_4$ resonance shown in Fig.\ref{figshort14} and in Fig.\ref{figevolc}. The high amplitude of the flipped spin $\sigma_z^{\rm{max}} \sim ~ 0.99$ is achieved for moderate and high charge relaxation times $T^{(2)} > 0.03 \mu s$ and for the long spin relaxation time $T^{(1)}$ when $T^{(2)}/T^{(1)} < 0.05$.} 
\label{figsigmafin}
\end{figure}

The predicted phenomenon of the locked flipped spin at the end of the evolution deserves a more detailed discussion. Namely, it is of interest to gather more information on the locked flipped spin amplitude as a function of the system parameters. First, we look at its dependence on the relaxation times which may vary depending on the structure type and quality. As before, we take equal energy and phase relaxation times for each subsystem, $T_\varphi^{(2)}=T_e^{(2)} \equiv T^{(2)}$, $T_\varphi^{(1)}=T_e^{(1)} \equiv T^{(1)}$. In Fig.\ref{figsigmafin} we plot the stabilized (i.e. obtained after a high number of driving periods) value of the spin projection $\langle \sigma_z \rangle_{t \to \infty}$ for different charge relaxation times $T^{(2)}$ labeled as differently shaped curves and covering a wide interval $(0.01 \ldots 0.15) ~\mu s$ which is consistent with the predictions for the GaAs-based double dots with the detuning \cite{Krzywda2021}. Each curve is plotted vs the ratio $T^{(2)}/T^{(1)}$ of the charge and spin relaxation time. All other parameters are the same as in Fig.\ref{figshort14}. This ratio is always lower than $1$ since the spin relaxation in our system is significantly slower than the charge relaxation. The other parameters are for the main harmonic of the $E_1-E_4$ resonance shown in Fig.\ref{figshort14} and in Fig.\ref{figevolc}. 

\begin{figure}[tbp]
\centering
\includegraphics[width=0.49\textwidth]{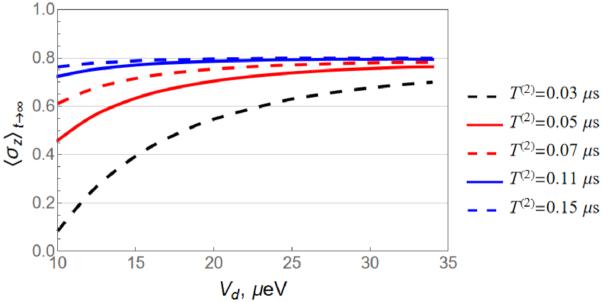}
\caption{Final value of the spin projection $\langle \sigma_z \rangle_{t \to \infty}$ as a function of the driving amplitude $V_d$ for different values of charge relaxation time $T^{(2)}$. The ratio $T^{(2)}/T^{(1)}=0.1$ and all other parameters are the same as in Fig.\ref{figsigmafin}. High amplitude of the locked flipped spin is achieved for the strong driving for all values of the charge relaxation time.} 
\label{figszfinvd}
\end{figure}

One can see in Fig.\ref{figsigmafin} that for a long spin relaxation time $T^{(1)}$ when the ratio $T^{(2)}/T^{(1)}$ is small enough the locked spin flip amplitude approaches high values for any curve corresponding to the different charge relaxation time $T^{(2)}$.  The highest amplitude of the flipped spin $\sigma_z^{\rm{max}} \sim 0.99$. We have already noted that the amplitude of the $z$ spin projection is smaller than $1$ in the presence of SOC so we believe that this result demonstrates the best available spin flip magnitude. 

Another dependence of the final amplitude of the flipped spin is shown in Fig.\ref{figszfinvd} as a function of the driving strength $V_d$. Different curves correspond to different charge relaxation times $T^{(2)}$ with the fixed ratio $T^{(2)}/T^{(1)}=0.1$ which is a typical value in the middle of Fig.\ref{figsigmafin} from which all other parameters are kept. According to Fig.\ref{figsigmafin}, for such ratio $T^{(2)}/T^{(1)}$ the best achievable amplitude of $\langle \sigma_z \rangle_{t \to \infty}$ is about $0.8$. It is evident from Fig.\ref{figszfinvd} that it can be achieved for a wide range of the charge relaxation time $T^{(2)}$ where all of the curves in the right side of Fig.\ref{figszfinvd} merge together.

\begin{figure}[tbp]
\centering
\includegraphics[width=0.49\textwidth]{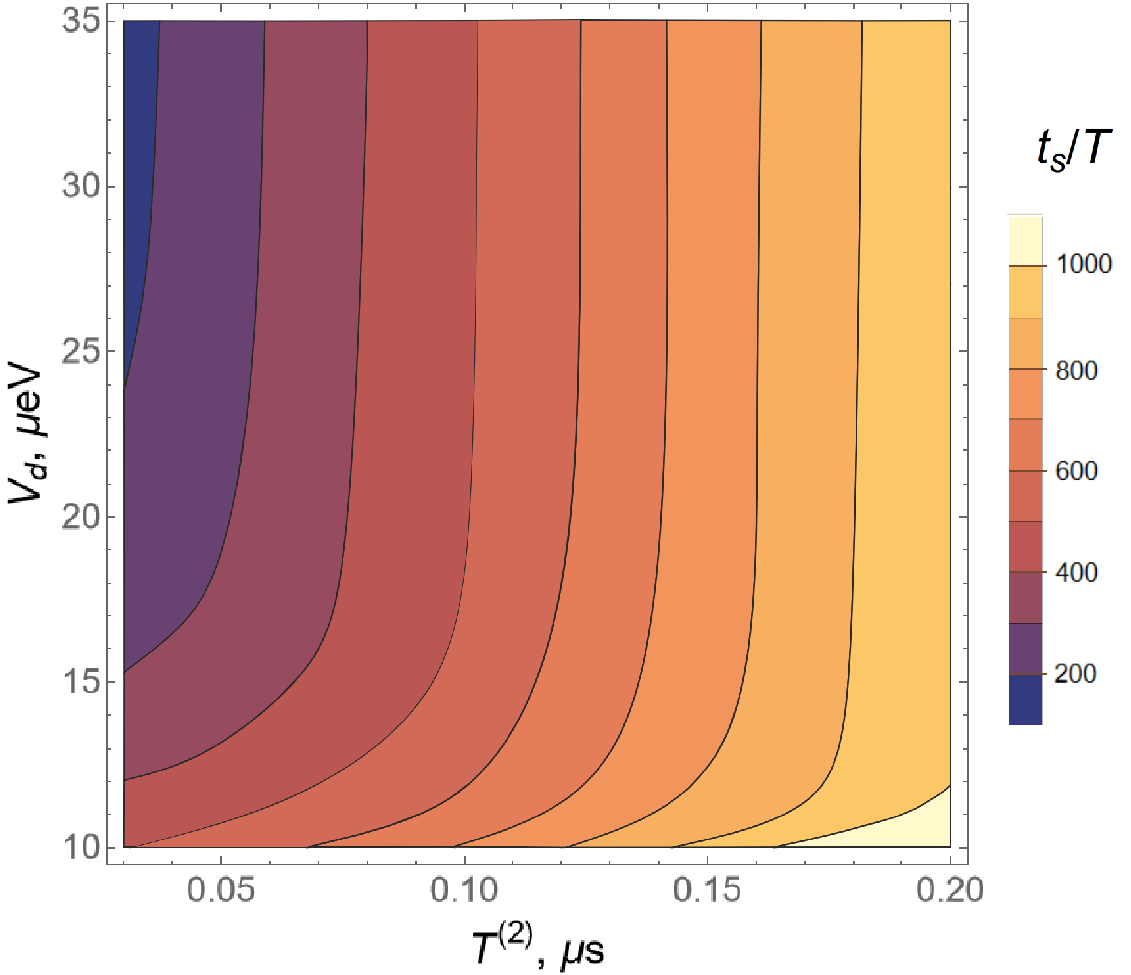}
\caption{Map of the spin flip stabilization time $t_s/T$ in units of the driving period $T=0.05$ $ns$ shown vs the charge relaxation time $T^{(2)}$ at fixed ratio $T^{(2)}/T^{(1)}=0.1$ and vs the driving amplitude $V_d$. All other parameters are the same as in Fig.\ref{figszfinvd}. For a long charge relaxation time (right part of the map) the dependence of $t_s/T$ on $V_d$ is weak and is determined mainly by the relaxation time itself. At shorter relaxation time (left part of the map) the dependence on $V_d$ is sharp, and the stabilized spin flip can be reached much faster.} 
\label{figfintime}
\end{figure}

Our final goal is to explore the dependence on the system parameters for the time when the stabilized spin flip amplitude is reached. In other words, we are interested to know not only how deep is the flip for the locked spin but also how fast this steady state can be reached. In Fig.\ref{figfintime} we show the map of the spin flip stabilization time $t_s$ defined as the characteristic time of the exponential fit $\sim (1- \exp (-t/t_s))$ for the envelope function fitted to the numerically obtained level population $|C_2(t)|^2$ in Fig.\ref{figevolc}. The time $t_s$ is measured in units of the driving period and is shown in Fig.\ref{figfintime} vs the charge relaxation time $T^{(2)}$ and the driving amplitude $V_d$. We believe that these two parameters are the most crucial for determining the $t_s$. All other parameters are the same as in Fig.\ref{figszfinvd} including the ratio $T^{(2)}/T^{(1)}=0.1$. Each point corresponds to the evolution tracked on a high enough number of the driving field periods $T=0.05$ $ns$ until the stabilized spin projection is achieved. One can see that for the slow relaxation corresponding to the long relaxation time at the right part of the map, the dependence of $t_s/T$ is rather shallow, and the obtained times are quite long. This can be explained as a long and coherent-like evolution in the case of slow relaxation where many spin oscillations are visible until the transition to the steady state is manifested, as it can be seen for the examples shown in Fig.\ref{fig14sh2} and Fig.\ref{figshort14} by the black and blue curves. As for the area of the faster relaxation in Fig.\ref{figfintime}, at the left part of the map, the dependence on the driving amplitude $V_d$ is much stronger, and the obtained stabilization time becomes significantly shorter if $V_d$ is increased. This is due to the faster damping of the coherent spin oscillations which dissipate quickly, unmasking the strong dependence of the spin flip time (the inverse of the Rabi frequency) on the driving field amplitude. To summarize, for the fixed ratio  $T^{(2)}/T^{(1)}$ we may interpolate the dependence of the $t_s$ in Fig.\ref{figfintime} by a simple formula
$t_s \approx a_{1}/ \Omega_R+ a_{2} T^{(2)}$ where $a_{1,2}$ are the fitting parameters. It can be easily verified that this approximation describes a correct topology of the contour plot curves in Fig.\ref{figfintime} if we remember that for the exact resonance the Rabi frequency $\Omega_R$ is proportional to the driving amplitude $V_d$. Another interesting feature of the results in Fig.\ref{figfintime} is that an increase in relaxation strength for the charge subsystem, i.e., the decreasing $T^{(2)}$ at a fixed ratio $T^{(2)}/T^{(1)}=0.1$ for a long-lived spin state, can play a positive role here, providing the stabilized spin flip to be reached faster. The stabilized spin flip state can be reached on the interval of several hundreds driving periods which for the parameters in Fig.\ref{figfintime} indicates the operation times of $15 \ldots 60$ $ns$ which is suitable for the possible information processing protocols.

To conclude, the observed effect of the spin flip lock for different relaxation times and for different driving amplitudes indicates that this effect can be found in various double dot setups in the gate-defined quantum dots, which makes it promising for the applications in spintronics and quantum computations.

\section{Conclusions}

We have studied the effects of relaxation on the spin and charge evolution in a driven double quantum dot modeling the spin and charge subsystems interacting via the spin-orbit coupling. The relaxation effects have been included via the density matrix formalism. The numerical analysis has demonstrated that the relaxation, leading to the trivial damping of the spin oscillations at the electric dipole spin resonance and its second subharmonic, plays a spin flip locking role for the multilevel dynamics where the spin flip is accompanied by an effective tunneling. The spin flip is found to be locked both on the main and on the neighboring subharmonic of the resonance, leading to the prolonged and stable spin flip in the desired quantum dot. The amplitude of the stabilized spin flip can reach as high as $0.99$ in the presence of spin-orbit coupling for a wide range of the relaxation parameters. The stabilized spin flip state can be reached on the interval of about $15 \ldots 60$ $ns$, making it suitable for the information processing. The predicted effect can be useful for the spin manipulation techniques in nanostructures and for the new coupled qubit proposals.

\section*{Acknowledgements}

The authors are grateful to S.A. Studenikin, Yu.G. Makhlin, V.A. Burdov and A.A. Konakov for fruitful discussions. The authors are supported by the Ministry of Science and Higher Education of the Russian Federation through the State Assignment No FSWR-2023-0035.


\begin{thebibliography}{99}

\bibitem{Nori2010}
S.N. Shevchenko, S. Ashhab, and F. Nori, Landau-Zener-
St{\"u}ckelberg interferometry, Phys. Rep. {\bf 492}, 1 (2010).

\bibitem{Nori2023}
O.V. Ivakhnenko, S.N. Shevchenko, F. Nori, Nonadiabatic
Landau-Zener-St{\"u}ckelberg-Majorana transitions, dynamics,
and interference, Phys. Rep. {\bf 995}, 1 (2023).

\bibitem{Semenov2007}
Y.G. Semenov and K.W. Kim, Elastic spin-relaxation processes in semiconductor quantum dots, Phys. Rev. B {\bf 75}, 195342 (2007).

\bibitem{Taylor2007}
J.M. Taylor, J.R. Petta, A.C. Johnson, A. Yacoby, C.M. Marcus, and M.D. Lukin, Relaxation, dephasing, and quantum control of electron spins in double quantum dots, Phys. Rev. B {\bf 76}, 035315 (2007).

\bibitem{Platero2008}
R. S{\'a}nchez, C. L{\'o}pez-Mon{\'i}s, and G. Platero, Coherent spin rotations in open driven double quantum dots, Phys. Rev. B {\bf 77}, 165312 (2008).

\bibitem{Petersson2010}
K.D. Petersson, J.R. Petta, H. Lu and A.C. Gossard, Quantum Coherence in a One-Electron Semiconductor Charge Qubit, Phys. Rev. Lett. {\bf 105}, 246804 (2010).

\bibitem{Ludwig2013}
P. Nalbach, J. Kn{\"o}rzer, and  S. Ludwig, Nonequilibrium Landau-Zener-Stueckelberg spectroscopy in a double quantum dot, Phys. Rev. B {\bf 87}, 165425 (2013).

\bibitem{Ribeiro2013}
H. Ribeiro, G. Burkard, J.R. Petta, H. Lu, and A.C. Gossard, Coherent Adiabatic Spin Control in the Presence of Charge Noise Using Tailored Pulses, Phys. Rev. Lett. {\bf 110}, 086804 (2013).

\bibitem{Vandersypen2013}
V. Srinivasa, K.C. Nowack, M. Shafiei, L.M.K. Vandersypen, and J.M. Taylor, Simultaneous Spin-Charge Relaxation in Double Quantum Dots,  Phys. Rev. Lett. {\bf 110}, 196803 (2013).

\bibitem{Vandersypen2014}
F.R. Braakman, J. Danon, L.R. Schreiber, W. Wegscheider, and L.M.K. Vandersypen, Dynamics of spin-flip photon-assisted tunneling, Phys. Rev. B {\bf 89}, 075417 (2014).

\bibitem{Loss2019}
T. Nakajima, A. Noiri, J. Yoneda, M.R. Delbecq, P. Stano, T. Otsuka, K. Takeda, S. Amaha, G. Allison, K. Kawasaki, A. Ludwig, A.D. Wieck, D. Loss and S. Tarucha, Quantum non-demolition measurement of an electron spin qubit, Nature Nanotech. {\bf 14}, 555 (2019).

\bibitem{Krzywda2021}
J.A. Krzywda and {\L}. Cywi{\' n}ski, Interplay of charge noise and coupling to phonons in adiabatic electron transfer between quantum dots, Phys. Rev. B {\bf 104}, 075439 (2021).

\bibitem{Platero2023}
D. Fern{\' a}ndez-Fern{\' a}ndez, J. Pic{\' o}-Cort{\' e}s, S. Vela Li{\~ n}{\'a}n and G. Platero, Photo-assisted spin transport in double quantum dots with spin–orbit interaction, J. Phys.: Materials {\bf 6}, 034004 (2023).

\bibitem{Loss2002}
H.-A. Engel and D. Loss, Single-spin dynamics and decoherence in a quantum dot via charge transport, Phys. Rev. B {\bf 65}, 195321 (2002).

\bibitem{Rashba2003}
L.S. Levitov and E.I. Rashba, Dynamical spin-electric coupling in a quantum dot, Phys. Rev. B {\bf 67}, 115324 (2003).

\bibitem{Loss2003}
D. Stepanenko, N.E. Bonesteel, D.P. DiVincenzo, G. Burkard, D. Loss, Spin-orbit coupling and time-reversal symmetry in quantum gates, Phys. Rev. B {\bf 68}, 115306 (2003).

\bibitem{Debald2005}
S. Debald and C. Emary, Spin-Orbit-Driven Coherent Oscillations in a Few-Electron Quantum Dot, Phys. Rev. Lett. {\bf 94}, 226803 (2005).

\bibitem{Loss2006}
V.N. Golovach, M. Borhani, and D. Loss, Electric-dipole-induced spin resonance in quantum dots, Phys. Rev. B {\bf 74}, 165319 (2006).

\bibitem{Yokoshi2009}
N. Yokoshi, H. Imamura, and H. Kosaka, Electrical Measurement of a Two-Electron Spin State in a Double Quantum Dot, Phys. Rev. Lett. {\bf 103}, 046806 (2009).

\bibitem{Nowack2007}
K.C. Nowack, F.H.L. Koppens, Yu.V. Nazarov, L.M.K. Vandersypen, Coherent Control of a Single Electron Spin with Electric Fields, Science {\bf 318}, 1430 (2007).

\bibitem{StehlikPetta2014}
J. Stehlik, M.D. Schroer, M.Z. Maialle, M.H. Degani, and J.R. Petta, Extreme Harmonic Generation in Electrically Driven Spin Resonance, Phys. Rev. Lett. {\bf 112}, 227601 (2014).

\bibitem{Studenikin2018}
A. Bogan, S. Studenikin, M. Korkusinski, L. Gaudreau, P. Zawadzki, A.S. Sachrajda,  L. Tracy, J. Reno, and T. Hargett, Landau-Zener-Stückelberg-Majorana Interferometry of a Single Hole, Phys. Rev. Lett. {\bf 120}, 207701  (2018).

\bibitem{Studenikin2019}
S. Studenikin, M. Korkusinski, M. Takahashi, J. Ducatel, A. Padawer-Blatt, A. Bogan, D. G. Austing, L. Gaudreau, P. Zawadzki, A. Sachrajda, Y. Hirayama, L. Tracy, J. Reno, and T. Hargett, Electrically tunable effective g-factor of a single hole in a lateral GaAs/AlGaAs quantum dot, Communications Physics {\bf 2}, 159 (2019).

\bibitem{Studenikin2023}
V. Marton, A. Sachrajda, M. Korkusinski, A. Bogan, and S. Studenikin, Coherence Characteristics of a GaAs Single Heavy-Hole Spin Qubit Using a Modified Single-Shot Latching Readout Technique, Nanomat. {\bf 13}, 950 (2023).

\bibitem{Petta2006}
J.R. Petta, A.C. Johnson, J.M. Taylor, A. Yacoby, M.D. Lukin, C.M. Marcus, M.P. Hanson, and A.C. Gossard, Charge and spin manipulation in a few-electron double dot, Physica E {\bf 34}, 42 (2006).

\bibitem{Burkard2007}
J. Kyriakidis and G. Burkard, Universal quantum computing with correlated spin-charge states, Phys. Rev. B {\bf 75}, 115324 (2007).

\bibitem{Zhang2007}
W.-M. Zhang, Y.-Z. Wu, C. Soo, and M. Feng, Charge-to-spin conversion of electron entanglement states and spin-interaction-free solid-state quantum computation, Phys. Rev. B {\bf 76}, 165311 (2007).

\bibitem{Ramon2010}
G. Ramon and X. Hu, Decoherence of spin qubits due to a nearby charge fluctuator in gate-defined double dots, Phys. Rev. B {\bf 81}, 045304 (2010).

\bibitem{Petersson2010b}
K.D. Petersson, C.G. Smith, D. Anderson, P. Atkinson, G.A.C. Jones, and D.A. Ritchie, Charge and Spin State Readout of a Double Quantum Dot Coupled to a Resonator, Nano Lett. {\bf 10}, 2789 (2010).

\bibitem{Mizuta2017}
R. Mizuta, R.M. Otxoa, A.C. Betz, and M.F. Gonzalez-Zalba, Quantum and tunneling capacitance in charge and spin qubits, Phys. Rev. B {\bf 95}, 045414 (2017).

\bibitem{Yang2019}
Y.-C. Yang, S.N. Coppersmith, and M. Friesen, High-fidelity single-qubit gates in a strongly driven quantum-dot hybrid qubit with 1/f charge noise, Phys. Rev. A {\bf 100}, 022337 (2019).

\bibitem{Kregar2020}
A. Kregar and A. Ram{\v s}ak, Rashba controlled two-electron spin-charge qubits as building blocks of a quantum computer, Mod. Phys. Lett. B {\bf 32}, 2040058 (2020). 

\bibitem{Rancic2022}
M.J. Ran{\v c}i{\' c}, Entangling spin and charge degrees of freedom in semiconductor quantum dots, Phys. Rev. A {\bf 105}, 032611 (2022).

\bibitem{Platero2022}
D. Fern{\'a}ndez-Fern{\'a}ndez, Y. Ban, and G. Platero, Quantum Control of Hole Spin Qubits in Double Quantum Dots, Phys. Rev. Appl. {\bf 18}, 054090 (2022).

\bibitem{PioroLadriere2007}
M. Pioro-Ladri{\'e}re, Y. Tokura, T. Obata, T. Kubo, and S. Tarucha, Micromagnets for coherent control of spin-charge qubit in lateral quantum dots, Appl. Phys. Lett. {\bf 90}, 024105 (2007).

\bibitem{Bodey2019}
J.H. Bodey, R. Stockill, E.V. Denning, D.A. Gangloff, G. {\' E}thier-Majcher, D.M. Jackson, E. Clarke, M. Hugues, C. Le Gall and M. Atat{\" u}re, Optical spin locking of a solid-state qubit, NPJ Quantum Information {\bf 5}, 95 (2019). 

\bibitem{Zuk2024}
I. Zuk, D. Cohen, A.V. Gorshkov, and A. Retzker, Robust gates with spin-locked superconducting qubits, Phys. Rev. Research {\bf 6}, 013217 (2024).

\bibitem{KS2022}
D.V. Khomitsky and S.A. Studenikin, Single-spin Landau-Zener-St{\"u}ckelberg-Majorana interferometry of Zeeman-split states with strong spin-orbit interaction in a double quantum dot, Phys. Rev. B {\bf 106}, 195414 (2022).

\bibitem{PRB2023}
D.V. Khomitsky, M.V. Bastrakova, V.O. Munyaev, N.A. Zaprudnov, and S.A. Studenikin, Controllable single-spin evolution at subharmonics of electric dipole spin resonance enhanced by four-level Landau-Zener-St{\"u}ckelberg-Majorana interference, Phys. Rev. B {\bf 108}, 205404 (2023).

\bibitem{BP}
H.-P. Breuer and F. Petruccione, {\it The theory of open quantum systems}, Oxford University Press (New York), 2003.

\bibitem{Hanggi1997}
S. Kohler, T. Dittrich, and P. H{\"a}nggi, Floquet-Markovian description of the parametrically driven, dissipative harmonic quantum oscillator, Phys. Rev. E {\bf 55}, 300 (1997).

\bibitem{GrifoniHanggi1998}
M. Grifoni, P. H{\"a}nggi, Driven quantum tunneling, Phys. Rep. {\bf 304}, 229 (1998).

\bibitem{Hone2009}
D.W. Hone, R. Ketzmerick, and W. Kohn, Statistical mechanics of Floquet systems: The pervasive problem of near degeneracies, Phys. Rev. E {\bf 79}, 051129 (2009).

\bibitem{Nalbach2009}
P. Nalbach and M. Thorwart, Landau-Zener Transitions in a Dissipative Environment: Numerically Exact Results, Phys. Rev. Lett. {\bf 103}, 220401 (2009).

\bibitem{Xu2014}
C. Xu, A. Poudel, and M.G. Vavilov, Nonadiabatic dynamics of a slowly driven dissipative two-level system, Phys. Rev. A {\bf 89}, 052102 (2014).

\bibitem{Makhlin2001}
Y. Makhlin, G. Sch\"on, and A. Shnirman, Quantum-state engineering with Josephson-junction devices, Rev. Mod. Phys. {\bf 73}, 357 (2001).

\bibitem{Satanin2012}
A.M. Satanin, M.V. Denisenko, S. Ashhab, and F. Nori, Amplitude spectroscopy of two coupled qubits, Phys. Rev. B {\bf 85}, 184524 (2012).

\bibitem{Bastrakova2021}
V.O. Munyaev and M.V. Bastrakova, Control of spectroscopic features of multiphoton transitions in two coupled qubits by driving fields, Phys. Rev. A {\bf 104}, 012613 (2021).

\bibitem{Pashin2023}
D.S. Pashin, P.V. Pikunov, M.V. Bastrakova, A.E. Schegolev,  N.V. Klenov, I.I. Soloviev, A bifunctional superconducting cell as flux qubit and neuron, Beilstein J. of Nanotechnol. {\bf 14}, 1116 (2023).

\bibitem{Bonifacio2020}
M. Bonifacio, D. Dom{\'i}nguez, and M.J. S{\'a}nchez, Landau-Zener-St{\"u}ckelberg interferometry in dissipative circuit quantum electrodynamics, Phys. Rev. B {\bf 101}, 245415 (2020).

\bibitem{Tokman2023}
M. Tokman, A. Behne, B. Torres, M. Erukhimova, Y. Wang, and A. Belyanin, Dissipation-driven formation of entangled dark states in strongly coupled inhomogeneous many-qubit systems in solid-state nanocavities, Phys. Rev. A {\bf 107}, 013721 (2023).

\bibitem{Marques2015}
B. Marques, A.A. Matoso, W.M. Pimenta, A.J. Guti{\' e}rrez-Esparza, M.F. Santos and S. P{\' a}dua, Experimental simulation of decoherence in photonics qudits, Sci. Rep. {\bf 5}, 16049 (2015). 

\bibitem{Zhang2023}
L. Zhang, L. Wang, M.F. Gelin, and Y. Zhao, Dynamics of dissipative Landau-Zener transitions in an anisotropic three-level system, J. Chem Phys. {\bf 158}, 204115 (2023).

\bibitem{Evers2005}
J. Evers, Phase-dependent interference mechanisms in a three-level
system driven by a quantized laser field, J. Mod. Opt. {\bf 52}, 2699 (2005).

\bibitem{Ilinova2012}
E. Ilinova and A. Derevianko, Dynamics of a three-level $\Lambda$-type system driven by trains of ultrashort laser pulses, Phys. Rev. A {\bf 86}, 013423 (2012).

\bibitem{Boukhris2023}
B. Boukhris, A. Tirbiyine, and J. El Qars, Gaussian quantum steering in a nondegenerate three-level laser, Mod. Phys. Lett. B {\bf 37}(35), 2350194 (2023).

\bibitem{Hofer2017}
P.P. Hofer, M. Perarnau-Llobet, L.D.M. Miranda, G. Haack, R. Silva, J. Bohr Brask, and N. Brunner, Markovian master equations for quantum thermal machines: local versus global approach, New J. Phys. {\bf 19}, 123037 (2017).

\bibitem{Cattaneo2019}
M. Cattaneo, G.L. Giorgi, S. Maniscalco, and R. Zambrini, Local versus global master equation with common and separate baths: superiority of the global approach in partial secular approximation, New J. Phys. {\bf 21}, 113045 (2019).

\bibitem{Lidar2016}
T. Albash, S. Boixo, D.A. Lidar, P. Zanardi, Quantum adiabatic Markovian master equations, New J Phys. {\bf 14}, 123016 (2012).

\bibitem{Petta2008}
J.R. Petta, J.M. Taylor, A.C. Johnson, A. Yacoby, M.D. Lukin, C.M. Marcus, M.P. Hanson, and A.C. Gossard, Dynamic Nuclear Polarization with Single Electron Spins, Phys. Rev. Lett. {\bf 100}, 067601 (2008).

\bibitem{Schuetz2014}
M.J.A. Schuetz, E.M. Kessler, L.M.K. Vandersypen, J.I. Cirac, and G. Giedke, Nuclear spin dynamics in double quantum dots: Multistability, dynamical polarization, criticality, and entanglement, Phys. Rev. B {\bf 89}, 195310 (2014).

\bibitem{Vezvaee2021}
A. Vezvaee, G. Sharma, S.E. Economou, and E. Barnes, Driven dynamics of a quantum dot electron spin coupled to a bath of higher-spin nuclei, Phys. Rev. B {\bf 103}, 235301 (2021).

\bibitem{Knap2013}
M. Knap, D.A. Abanin, and E. Demler, Dissipative Dynamics of a Driven Quantum Spin Coupled to a Bath of Ultracold Fermions, Phys. Rev. Lett. {\bf 111}, 265302 (2013).

\bibitem{DasSarma2016}
S. Das Sarma, R.E. Throckmorton, and Y.-L. Wu, Dynamics of two coupled semiconductor spin qubits in a noisy environment, Phys. Rev. B {\bf 94}, 045435 (2016).

\end{thebibliography}
\end{document}